\documentclass[manuscript]{aastex}

\usepackage{graphicx}
\usepackage{xcolor}
\usepackage{multirow}

\newcommand{\Lsun}{\mbox{$log_{10}(L/L_{\odot}$})}
\newcommand{\Rsun}{\mbox{$R/R_{\odot}$}}

\shorttitle{Parallaxes of five L Dwarfs with a Robotic Telescope}
\shortauthors{Wang et al.}
\begin{document}

\title{Parallaxes of five L Dwarfs with a Robotic Telescope}

\author{Y. Wang\altaffilmark{1,2,3},\email{youfenwang@shao.ac.cn}
  H.R.A. Jones\altaffilmark{3},\email{h.r.a.jones@herts.ac.uk} R.L. Smart\altaffilmark{4},\email{smart@oato.inaf.it} F. Marocco\altaffilmark{3},\email{federico.marocco@gmail.com} D.J. Pinfield\altaffilmark{3},\email{d.j.pinfield@herts.ac.uk}
   Z. Shao\altaffilmark{1,10},\email{zyshao@shao.ac.cn} I.A. Steele\altaffilmark{5}, \email{ias@astro.livjm.ac.uk} Z. Zhang\altaffilmark{3},\email{z.zhang7@herts.ac.uk} A.H. Andrei\altaffilmark{4,6},\email{oat1@on.br} 
  A.J. Burgasser\altaffilmark{9},\email{ajb@mit.edu} 
  K.L. Cruz\altaffilmark{7,8},\email{kelle.cruz@hunter.cuny.edu} J. Yu\altaffilmark{1,2},\email{yujc@shao.ac.cn} 
  J.R.A. Clarke\altaffilmark{11},\email{jraclarke@hotmail.com} C.J. Leigh\altaffilmark{5},\email{cjl@astro.livjm.ac.uk} 
  A. Sozzetti\altaffilmark{4},\email{sozzetti@oato.inaf.it} D.N. Murray\altaffilmark{3},\email{D.Murray@herts.ac.uk} \&
  B. Burningham\altaffilmark{3},\email{b.burningham@herts.ac.uk}}

\altaffiltext{1}{Shanghai Astronomical Observatory, Chinese Academy of Sciences, 80 Nandan Road, Shanghai, 200030, China.}
\altaffiltext{2}{University of Chinese Academy of Science, 19 Yuquan Road, Shijingshan District, Beijing, 100049, China}
\altaffiltext{3}{Center for Astrophysics Research, University of Hertfordshire, Hatfield AL10 9AB,UK}
\altaffiltext{4}{INAF/Osservatorio Astrofisico di Torino, Strada Osservatorio 20, 10025 Pino Torinese, Italy}
\altaffiltext{5}{Astrophysics Research Institute, Liverpool John Moores University, Liverpool, CH41 1LD, UK}
\altaffiltext{6}{Observat\'{o}rio Nacional/MCT,R, Gal. Jos\'{e} Cristino 77,CEP20921-400,RJ, Brazil }
\altaffiltext{7}{Dept. of Astrophysics, American Museum of Natural History, 79th St at Central Park West, New York, NY 10024}
\altaffiltext{8}{Dept. of Physics and Astronomy, Hunter College, City University of New York, 695 Park Avenue, New York, NY 10065}
\altaffiltext{9}{Center for Astrophysics and Space Science, University of California San Diego, La Jolla, CA 92093}
\altaffiltext{10}{Key Laboratory for Astrophysics, Shanghai 200234, China}
\altaffiltext{11}{Departamento de F\'{i}sicay Astronom\'{i}a, Facultad de Ciencias, Universidad de Valpara\'{i}so, Av. Gran Breta\~na 1111, Playa Ancha,
                   Casilla 5030, Valpara\'iso, Chile}

\begin{abstract}
  We report the parallax and proper motion of five L dwarfs obtained with
  observations from the robotic Liverpool Telescope. Our derived proper
  motions are consistent with published values and have considerably
  smaller errors. Based on our spectral type versus absolute magnitude
  diagram, we do not find any evidence for binaries among our sample, or, at
  least no comparable mass binaries. Their space velocities locate them within
  the thin disk and based on the model comparisons they have solar-like
  abundances. For all five objects, we derived effective temperature,
  luminosity, radius, gravity and mass from a evolutionary model(CBA00) and our
  measured parallax; for three of the objects, we derived their effective
  temperature by integrating observed optical and near-infrared spectra and
  model spectra(BSH06 or BT-Dusty respectively)
  at longer wavelengths to obtain bolometric flux and then using the classical
  Stefan-Boltzmann law: generally the three temperatures for one object derived
  using two different methods with three models are consistent, while at lower
  temperature(e.g. for L3) the differences among the three temperatures
  are slightly larger than that at higher temperature(e.g. for L1).
\end{abstract}

\keywords{stars: brown dwarfs, parallax, proper motion --- Data Analysis and Techniques}

\section{Introduction}
L-type dwarfs are ultracool objects cooler than M dwarfs. Most L dwarfs are
expected to be brown dwarfs, i.e., have insufficient mass to achieve the
central temperatures and pressures necessary for {sustained} hydrogen
burning. Brown dwarfs have physical properties intermediate between the least
massive stars and the most massive planets and are thus a useful bridge
between studies of stars and planets \citep{bur11}. However, the lack of a
unique age - mass - spectral type relationship leads to distance being a
critical parameter to understand brown dwarfs. A distance is required to
derive an absolute magnitude and hence energy output. Parallaxes are a model
independent parameter that can be used to constrain radius or temperature thus
allowing modeling to explore relations between other parameters, mass -
surface gravity- age - metallicity, more freely. Considering that distances
are so valuable, it is a sign of the difficulty in obtaining them that out of
more than {900} known L dwarfs (www.dwarfarchives.org hereafter DwarfArchive)
less than 90 have measured parallaxes and, when this programme started, there
were less than 20.

Here we discuss the determination of parallax and proper motion for five L
dwarfs using the robotic Liverpool Telescope\footnote{telescope.livjm.ac.uk}
(hereafter LT, Steele et al. 2004). In general, the observations required for
parallax determinations are quite simple and routine. The important
characteristics for observations in a parallax program are stability in the
instrumental setup and repeatability in the observational procedure.  The
rigorous scheduling criteria, efficient use of time, flexibility in
scheduling, and, observational consistency in robotic observations make it a
very attractive possibility for parallax programs. This program was
envisioned to see if the LT could become an exemplar parallax machine for
future parallaxes of bright brown dwarfs and nearby red dwarfs. The number of
brown dwarfs is increasing rapidly with continued discoveries from wide-field 
sky survey program, e.g., SDSS
\citep{yor00}, VISTA \citep{eme01}, CFHT \citep{mon07}, UKIRT \citep{law07},
and WISE \citep{wri10} surveys. Many of these are
observable with the LT.

This paper is divided into seven sections. In Section 2 we describe the
observations and data reduction procedures. In Section 3 we report the main
astrometric results. In Section 4 we study the binarity, Galactic membership,
metallicity and gravity properties using spectral type - absolute magnitude,
U-V velocity and color- absolute magnitude with model tracks diagrams.  In
Section 5 we present the bolometric flux, luminosity and effective temperature
of our targets obtained combining our measured parallaxes with the
optical/infrared spectra and evolutionary/atmospheric models. In Section 6 we comment on
individual objects and in Section 7 we summarize our findings and briefly
  describe our future work plan.

\section{Observations and reduction procedures}

The LT is a totally robotic telescope located at the Observatorio del Roque de
Los Muchachos on the Canary island of La Palma in Spain and operated by the
Liverpool John Moores University in the United Kingdom. It is an Alt-Az
telescope with Ritchey-Chretien Cassegrain optics with a primary mirror of 2.0
m. In 2004, when this parallax program started, there were two instruments
that were suitable for a brown dwarf parallax program: SupIRCam and
RATCam. SupIRCam is an infrared sensitive 256x256 pixel HgCdTe array with a
pixel scale of 0.413 {\rm arcsecond/pixel} and a field of view of 1.7 {\rm
  arcmin}. RATCam is an optically sensitive 2048x2048 pixel CCD camera with a
pixel scale of 0.1395 {\rm arcsecond/pixel} and a field of view of 4.6 {\rm
  arcmin}. The SDSS-$z$ filter (hereafter simply $z$) corresponds to the
brightest optical magnitude for L dwarfs. The larger field, smaller pixel
scale and similar required exposure times for typical L dwarfs of the RATCam
instrument in the $z$ band filter compared to the SupIRCam in the J filter led
to RATCam being the preferred choice for our program.

\subsection{Targets}

The target list was selected from the literature with the following criteria:
visible to the LT, a $z$ band magnitude brighter than 18, no published
trigonometric parallax in 2004 and those objects with the smallest photometric
distance were preferred.  Here we report on the five that have enough
observations spread over 2004/2005, 2008/2009 and 2011/2012 to provide
reliable parallaxes. In Table \ref{WPit1} we list the five objects with their
$z$, and estimated $z$, 2MASS $JHK$ \citep{skr06} and WISE $W1$ \citep{wri10}
band magnitudes, optical and near-infrared spectral types.

\subsection{Observational procedure}
The five targets were observed between August 2004 and July 2012 using RATCam
with the $z$ band filter. In order to minimize the effect of differential
color refraction, we observe when the targets are within 30 {\rm minutes} of
the meridian. The observations were primarily made during twilight hours,
since this is when the objects have maximum parallax factors in right
ascension.  Observed this way, the data are primarily located on the ends of
the major semi-axis of their parallax ellipse. During each observation we take
three exposures of 160{\rm s} to allow for robust removal of cosmic rays and
to diminish the random errors. One exposure of this length nominally provides
a signal-to-noise of more than 50 on these targets.

Differential color refraction \citep[DCR,][]{mon92, sto02} is the small
varying displacement of objects with different colors in a field that results
from the variation of the atmosphere refractive index with wavelength. It is
strongest in the blue bands and gradually gets very small in the infrared. The
targets in this parallax program are redder than the anonymous reference
objects so this displacement is systematically different from the average of
those reference objects. We request that all our observations are made within
30 minutes of the meridian so the variation in airmass, and hence differential
movement, is minimized. In the Torino Observatory Parallax Program (hereafter
TOPP, Smart et al. 1999) we found the effect was very small in the I band and
it will be smaller for the $z$ band, though L dwarfs are redder than the TOPP
targets. In the work by Dahn et al. (2002) \nocite{dah02} they do not include
DCR terms as they found they changed the $z$ band parallax of L and T objects
by only 0.3 mas. In Albert et al. (2011) \nocite{alb11} they also found the DCR
in the $z$ band on relative astrometry of brown dwarfs was small enough to
neglect. Following these results, and in light of our observational criteria,
we have not included DCR terms in this analysis.  For future work we will
carry out a number of experiments to measure the DCR in the LT $z$ band system
and review this decision.

\subsection{Reduction Procedures}

The bias subtraction, trimming of the overscan regions, dark subtraction and
flat fielding are carried out via the standard LT pipeline
\citep{ian04}. However, images in the $z$ band display prominent fringes
caused by thin-film interference \citep[see Appendix A in][]{ber08}. Fringes
have a small effect on the photometry and astrometry for bright objects, but
can have a significance impact for faint objects when their fluxes are
comparable with the intensity of the fringes. Since our targets are relatively
faint, we must investigate the impact of fringes.

To examine the intensity and evolution of the fringes we divide the images
into three sub-samples: (1) 2004/2005, (2) 2008/2009 and (3) 2011/2012
images. Each sub-sample contains several hundreds of frames.  For each frame
in each sub-sample we pick out an empty area (of 100$\times$100~pixels) which
is seriously fringed but without any or with very few objects. We calculate
the RMS of the counts and average the values within each sub-sample. We find
the count variations before and after defringing are {11.2 and 9.5, 8.1 and 6.8,
8.6 and 7.5} respectively for the three subsets, a difference that we consider
significant.

The standard LT pipeline constructed biannual fringe maps and our first
attempt was to use the most appropriate for each night. However, fringes are
dependent on the sky conditions at the time of observation and vary during the
course of a night. The ideal case would be to make a fringe map for each
image, but this is not feasible. In addition we usually only have a few images
in any given night so even a per night fringe map is not possible. Our second
attempt was to construct fringe maps following the recipe in Andrei et
al. (2011) for subsets of 20-30 frames while attempting to keep nights and
periods covered intact. Using the fringe maps constructed by ourselves usually
gave similar parallax results to those using the LT fringe map except in the
case of the fainter targets. This is probably due to the fact that sometimes
to have sufficient frames to construct a fringe map we had to include a
relatively long time-span but with few frames compared to the LT fringe
maps. The results presented here used the LT fringe maps which also produced
more robust parallax solutions.

\subsection{Centroid Precision}
Since our targets are faint and our data impacted by fringes which we can not
remove completely, it is critical to have appropriate centroiding software in
order to determine their position which is the fundamental data for a parallax
determination. We tried several different methods: (1) two-dimensional
Gaussian fit to the point spread function as used in the TOPP, (2) the widely
used Sextractor routine which is designed for large scale galaxy surveys and
also works well on moderately crowded star fields
(http://www.astromatic.net/software/sextractor) and (3) the maximum likelihood
barycenter as implemented in the imcore software of the Cambridge Astronomy
Survey Unit (CASU,
http://casu.ast.cam.ac.uk/surveys-projects/software-release).

\begin{figure}[!htp]
\includegraphics[width=3.in]{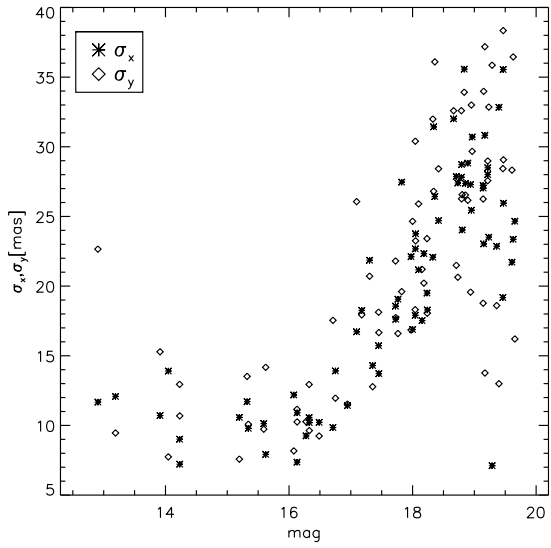}
\caption{$\sigma_{x}$,$\sigma_{y}$ of the {\em x,y}
  coordinates for common objects in 57 frames of the
  2M1807+5015 field. The frames were made over 19
  nights spanning $\sim$7.04 years. On the x axis we
 plot apparent magnitude in the $z$ band and on the y
 axis we plot the $\sigma_{x}$,$\sigma_{y}$ in {\rm mas}.
 The median $\sigma_{x}$,$\sigma_{y}$ are 21,21
 {\rm mas} for all objects and 11,11 {\rm mas} for
  objects with $z$ band magnitude brighter than 17.}
\label{WPif1}
\end{figure}

We tested all the centroiding procedures by comparing object positions from 57
frames of the 2M1807+5015 field. The centroiding was also tested with
different defringing procedures. We found that for brighter objects we get
similar results but CASU imcore centroids work significantly better for the
fainter ones giving smaller errors.  If we do not defringe, the median
$\sigma_{x}$ ,$\sigma_{y}$ {for} the {\em x,y} coordinates are 25,28 {\rm mas}
for all objects, and 13,14 {\rm mas} for objects brighter than magnitude
z=17. Applying the fringe map provided by LT pipeline we find that the
precision improves to 21,21 {\rm mas} and 11,11 {\rm mas} respectively. In
Fig. \ref{WPif1} we present the standard deviations of the object
coordinates in the 2M1807+5015 sequence defringed with the LT biannual fringe
maps and centroided with the CASU routines.

Based on our experience in other parallax programs we expected to achieve a
lower floor than 11,11 {\rm mas} for the centroiding precision. 
We note that the RATCam CCD has electronic gates aligned with x axis and 
physical gates aligned with y axis. The precision from electronic gates 
is better than the physical gates. The source of
this higher noise is probably because that nominally $x$ is orientated in the
direction of RA and $y$ in Dec, but due to flexibility problems with the RATCam
coolant pipes it was not possible to always keep the same alignment. A
procedure of "cardinal pointing" is adopted that aligns the rotator to one of
the four cardinal positions: 0, 90, 180 and 270. A third of our observations
have the rotator aligned to 0, that is with North at top, East at left.  The
other images are evenly distributed between the other cardinal points, except
during the first year of observation when there are also non-standard
positions with a number of different angles. Since the astrometric distortion
is partially a function of the focal plane variations, this physical rotation
of the focal plane impacts negatively on the expected precision. The new
Infrared-Optical camera on the LT does not have this constraint.

\begin{figure}[!htp]
  \includegraphics[width=3.in]{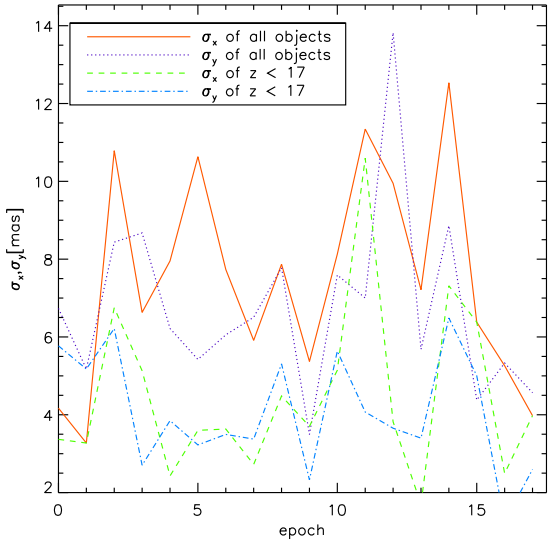}
  \caption{$\sigma_{x}$,$\sigma_{y}$ of the {\em x,y} coordinates for objects in
 the 2M1807+5015 field as a function of epoch sapnning 7.04 years.
 The $\sigma_{x}$,$\sigma_{y}$ are calculated using three sequential images
 from each of the 19 nights. The $\sigma_{x}$,$\sigma_{y}$ along the y axis are
 median value of the corresponding epochs for two subsets: (1) all objects
 detected and (2) objects detected brighter than 17 in $z$ magnitude. The
 median precisions among the 19 epochs for all objects are 7.7,6.5 mas and for
 the $z <$ 17 subset are 3.8,3.7 mas in $\sigma_{x}$,$\sigma_{y}$
 respectively.}
 \label{WPif2}
\end{figure}

Another possible source for this high floor is that our observations for
2M1807+5015 covered several years and there will be a small contribution from
random proper and parallactic motion of the reference stars. Since we expose
three times for each target in each night the precision using these three
frames excludes this random motion contribution.  There are 3 observations on
19 nights so we have 19 sub-groups with 3 frames in each. For each sub-group we
calculated their median $\sigma_{x}$,$\sigma_{y}$ for all objects and for
objects brighter than $z$=17. In Fig \ref{WPif2} we plot the sigma versus
epoch, the median precision for the objects brighter than $z$=17 improved to
3.8,3.7 mas.  As each sample comprises of only three images we expect this to
be an underestimate of the true sigma but it supports our hypothesis of the
contribution from random motions. Since our parallax solutions come from the
combined data-sets we must include the instrumental and reference system
variations, so considering consistency of the final and per-epoch errors we
assume an observational precision of 11,11 {\rm mas}.

\section{Parallaxes and Proper Motions}

Using the {\em x,y} coordinates determined from the CASU imcore software we
derived the parallaxes and proper motions using the methods adopted in the
TOPP (Smart et al. 2003, 2007).  The software selects the frames and reference
stars automatically, for example frames with less than four reference stars in
common, or, stars with large errors or high proper motions, are dropped. A
base frame is selected in the middle of the sequence with a high number of
stars. This base frame is transferred to a standard coordinate system using
objects in common with 2MASS. The other frames are then transferred to this
system using all common stars with a linear transformation.  Then by fitting
the combined observations of the target in the standard coordinate system we
find its relative parallax and proper motion. The correction from relative to
absolute parallax is calculated using the galaxy model of \citet{men96} as
described in Smart et al. (2003).  We estimate the error on this correction to
be around 30\% or 0.4-0.6 mas for these fields \citep{sma07}, which is
negligible compared to the formal error of the parallaxes.

\begin{figure}[!htp]
\includegraphics[width=3.in]{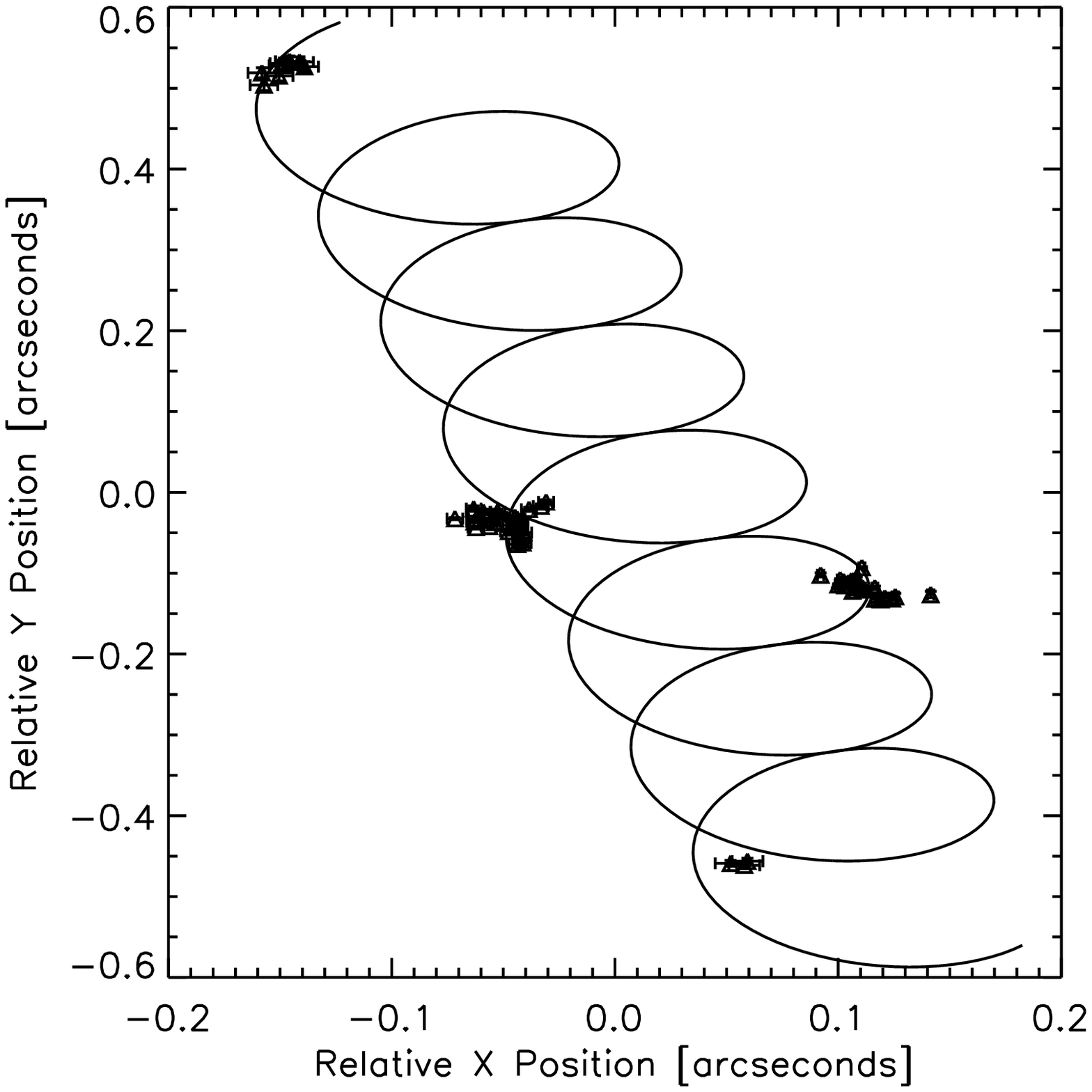}
\includegraphics[width=3.in]{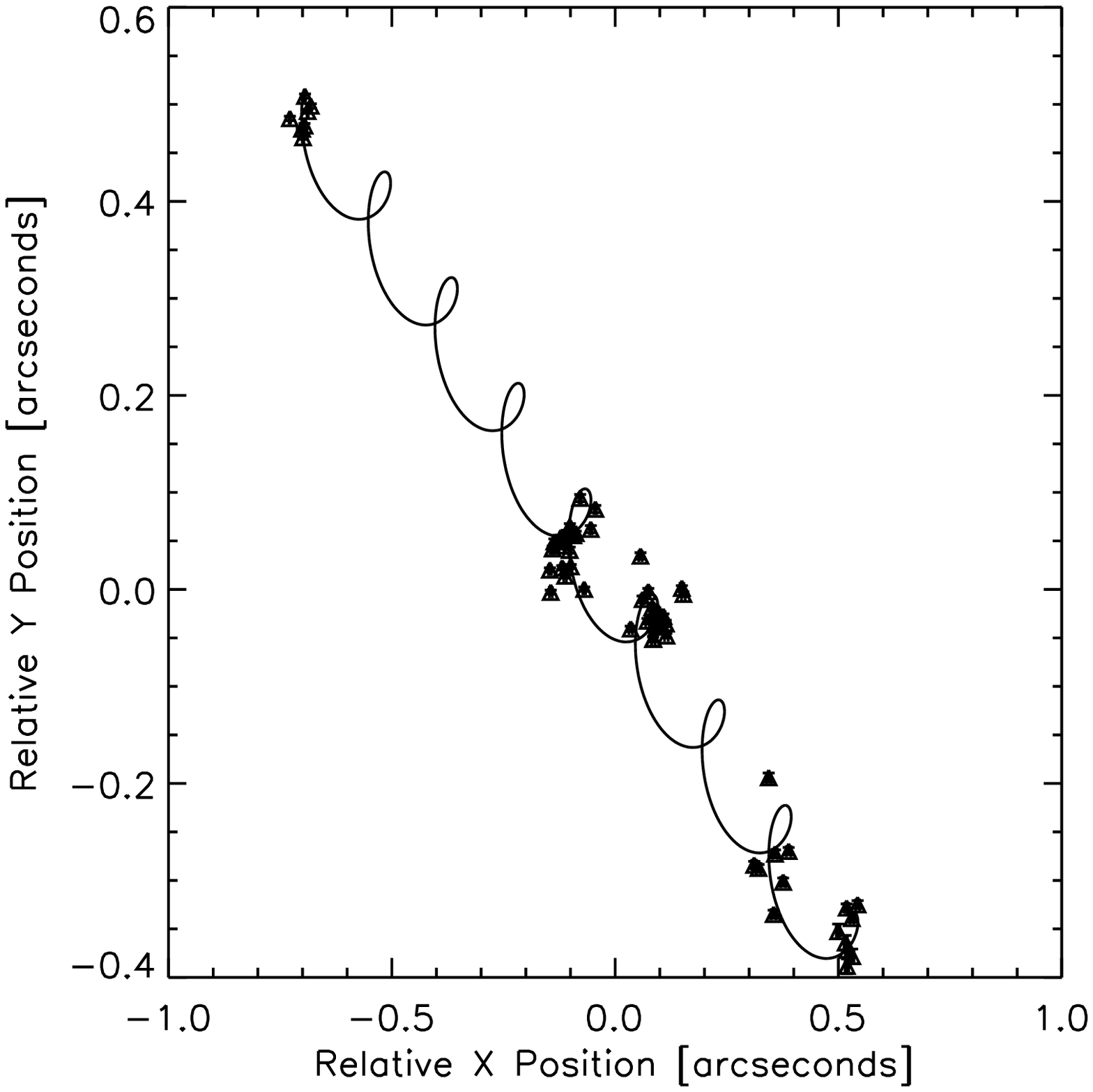}
\caption{Observations of 2M1807+5015 (left) and 2M2238+4353 (right) using CASU
  centroids along with our solutions over plotted. }
\label{WPif3}
\end{figure}

In Table \ref {WPit2} we list our results and in Fig. \ref{WPif3} we plot
the solutions for the targets 2M1807+5015 and SD1717+6526 which have
respectively the lowest and highest parallax errors. As shown in Fig.
\ref{WPif1} the centroiding deteriorates significantly as the object gets
fainter. This is reflected in the correlation of derived parallax precision
with apparent magnitude in Tables \ref{WPit1} and \ref{WPit2} and explains the
noisier observations of SD1717+6526.


\section{Analysis of properties}
In this section we examine the physical characteristics of
our objects using our parallax and proper motion results and
 taking advantage of two different brown dwarf models.

\subsection{Absolute magnitude and spectral type interpretation}
\begin{figure}[!htp]
\includegraphics[width=3.in]{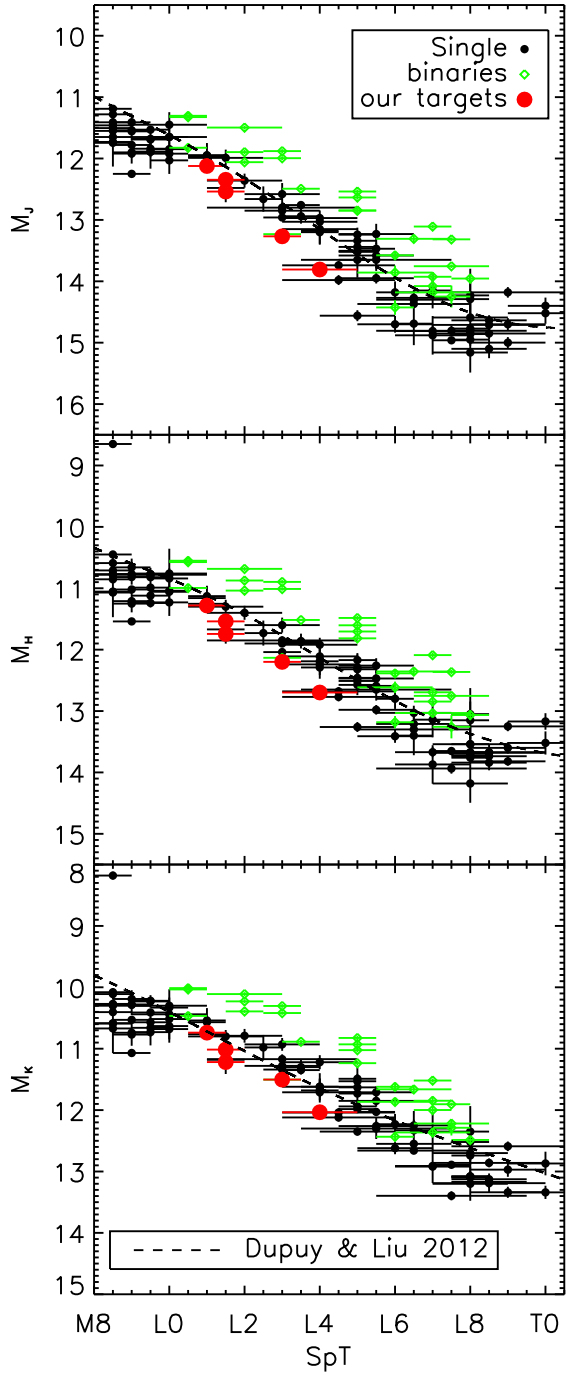}
\caption{{\rm 2MASS} JHK absolute magnitude as a function of optical spectral
  type.  The black solid circles and the blue diamonds are objects from Dupuy \& Liu (2012)
  with published parallaxes. Black solid circles are M8.5 to T0 dwarfs without
  indication of binarity, blue diamonds are unresolved binaries. The red triangls
  are our five targets. Including our five targets, in total 84 single objects
  are used when fitting the solid red polynomial curve. Dupuy \& Liu (2012)
  relations are over plotted as dashed lines.}
\label{WPif4}
\end{figure}

In Fig. \ref{WPif4} we plot the optical spectral type versus absolute
magnitude diagram in J, H and K bands including our objects and published
objects with measured parallaxes from Dupuy \& Liu (2012).
The solid red lines are the polynomial fit from M8 to T0 including our five L
dwarfs and the published objects but excluding known and possible binaries.
The magnitudes are 2MASS values and the spectral types are from optical
spectra.  We also include the Dupuy \& Liu (2012) spectral type versus
absolute magnitude relation and note that our targets and fit (solid red
lines) are slightly below their relation (dashed lines). In Table \ref{WPit3} we
compare the Dupuy \& Liu (2012) spectrophotometric distances with our trigonometric
ones, and the two distances for the five L dwarfs are generally consistent within one sigma.
 Our trigonometric distance are generally slightly smaller,
however, with such a small sample it is not possible to draw any conclusions.

In Table \ref{WPit5} we list the coefficients and errors of the fit to the
polynomial:
\begin{equation}
M_{X}= \sum_{i=0}^{6}a_{i}x(SpT)^{i}
\label{WPie1}
\end{equation}
where SpT indicates the spectral type, following the convention M0 = 0, ...L0
= 10, ... T0 = 20, and $M{_X}$ is the absolute magnitude in the X band where
{\it X}=2MASS J, H or K. The fit is valid only in the SpT range from M8
to T0.

If any of our targets are unresolved binaries they will be brighter than a
single object in Fig. \ref{WPif4}. This brightening reaches a maximum for
equal-mass binaries with an expected difference of 0.75 {\rm mag}. Since this
is not the case, we conclude none of the five targets consist of comparable
mass binaries.

\subsection{Kinematic analysis}

The velocities of nearby objects are dominated by their rotation around the
Galactic center. But they also have peculiar motions of several tens of
km/s. In the Galactic coordinate system, this spatial motion can be described
using $U,V$ and $W$ velocities, with the $U$ axis oriented towards the Galactic
anti-center. Different stellar populations such as disk objects or halo
objects have particular distributions in $U,V$ and $W$ velocity space. So, if we can
obtain the $U, V$ and $W$ of an object, we can kinematically determine which
Galactic component it belongs to.

We convert proper motion, parallax and radial velocity into $U, V$ and $W$ velocities
 listed in Table \ref{WPit6}. All velocities are corrected
to the LSR adopting the solar motion $U_{\odot}$, $V_{\odot}$, $W_{\odot}$ =
11.10, 12.24, 7.25 \citep{sch10}. Only two objects, 2M0141+1804 and
2M1807+5015, have published radial velocity data ($24.7$ and $-0.4$ km/s),
obtained from high resolution spectroscopy \citep{bla10}. For the other
objects we assume their radial velocity distribution is similar to M dwarfs
and we can then estimate their membership statistically as described below.

We select 18563 nearby M dwarfs within 500 {\rm pc} from the \citet{wes11}
spectroscopic catalog of $\sim 70000$ M dwarfs from the SDSS DR7 which have
measured radial velocity and spectrophotometric distances for each star. 
The radial velocity distribution of this sub-sample follows a Gaussian 
profile with the mean velocity $\sim 0$ km/s and $\sigma$ $\sim 30$ km/s.

\begin{figure}[!htp]
\includegraphics[width=3.in]{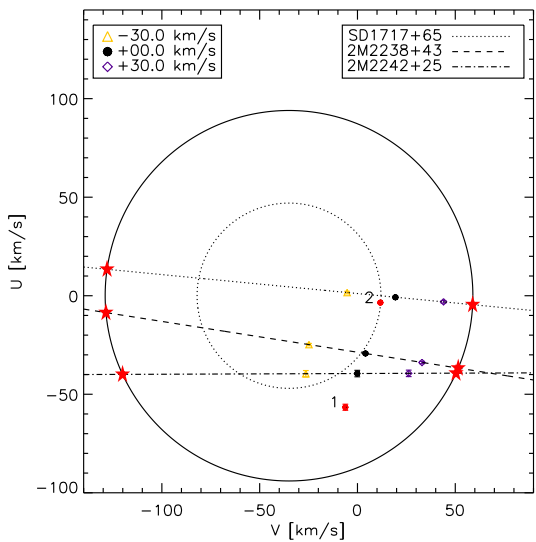}
\caption{$U$ versus $V$ Galactic velocities. The dotted and solid ellipses
     are 1 and 2$\sigma$ velocity ellipsoid for disk sars according to
     \citet{rei01}. The coordinates of the center are  (-35,0) {\rm km/s}
    \citep{opp01}; the radius is 47 {\rm km/s} and 94 {\rm km/s}. The red
   filled circles labeled 1 \& 2 are 2M0141+1804 and 2M1807+5015 which have
   measured radial velocities. For the other objects the dotted, dashed
   dot and dashed lines describe the $U$ and $V$ velocities when adopting different
   radial velocities. Their $U$ and $V$ velocities when using 1$\sigma$ and mean
   radial velocity 30, 0, -30 {\rm km/s} are shown on each line. The asterisks
   located on the 2 sigma ellipse indicate U and V velocities from the "critical"
   radial velocities listed in Table \ref{WPit7}. }
\label{WPif5}
\end{figure}

We use a check from \citet{opp01} to identify their membership in the
different Galactic components.  Objects that satisfy $[U^{2}+(V+35)^{2}]^{1/2}
> 94$ km/s are considered halo objects at the $2\sigma$ level. In the U-V
diagram Fig.\ref{WPif5}, we find 2M0141+1804 and 2M1807+5015 both locate
within the $2\sigma$ circle. Which means the two targets are probably disk
objects. We assume that the other three L dwarfs without  measured radial velocity
 have velocities which follow the Gaussian distribution of the SDSS
M dwarfs' as described above and then plot their $U$ and $V$ projection along the
straight lines as shown in Fig.\ref{WPif5}. On these lines we plot three
points for each object, indicating their $U$ and $V$s when adopting radial velocity 0
and $\pm$30 km/s. Since all space velocities are located within the $2\sigma$
circle, it is likely that these three L dwarfs are disk objects.

To further quantify the possibility of the three L dwarfs without measured radial
velocity being halo component, we calculate two "critical" radial velocities 
(expressed as $V_{\rm rad1}$ and $V_{\rm rad2}$) which locate the $ U$ and $V$ velocities on the
$2\sigma$ circle. Considering their radial velocity distribution,
integrate the Gaussian profile of the radial velocity outside the two
points $V_{\rm rad1}$ and $V_{\rm rad2}$, we then obtained probabilities for the
three objects being halo component, which are listed in Table
\ref{WPit7}. These three objects have small probabilities of being halo
component. So we conclude that our five L dwarfs are likely disk objects.

We use the test from Section 5 of \citet{mar10} to see if these objects are
very young.  Younger stars have a small space velocity dispersion and hence
small space velocities. Objects with U between -20 and 50 km/s, V between -30
and 0 km/s , W between -25 and 10 km/s will be younger than 0.5 Gyr. With the
U,V and W ranges presented in Table \ref{WPit6} it is unlikely that these objects
are younger than 0.5 Gyr.

\subsection{Comparison with model predictions}

\begin{figure}[!htp]
\includegraphics[width=3.in]{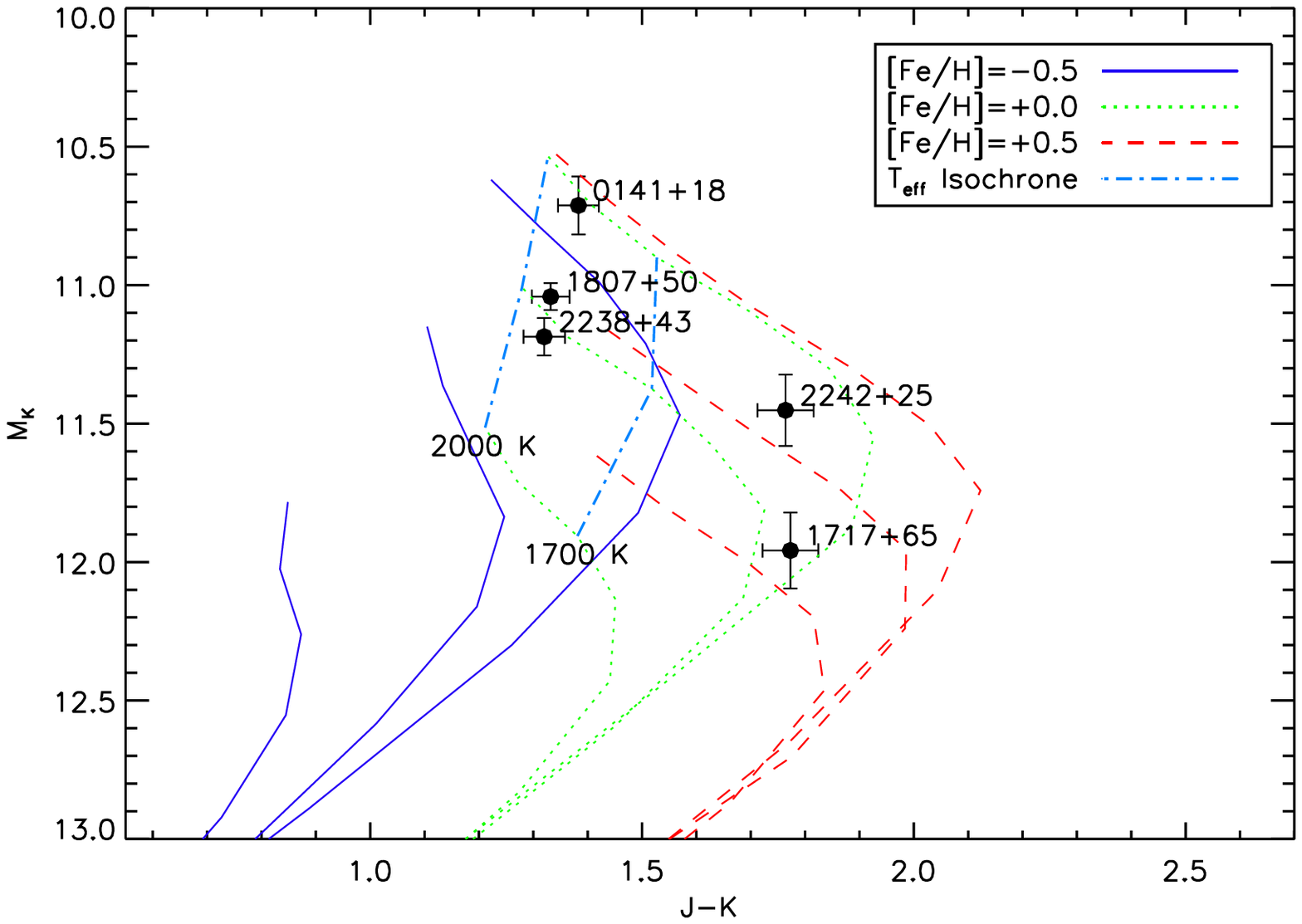}
\includegraphics[width=3.in]{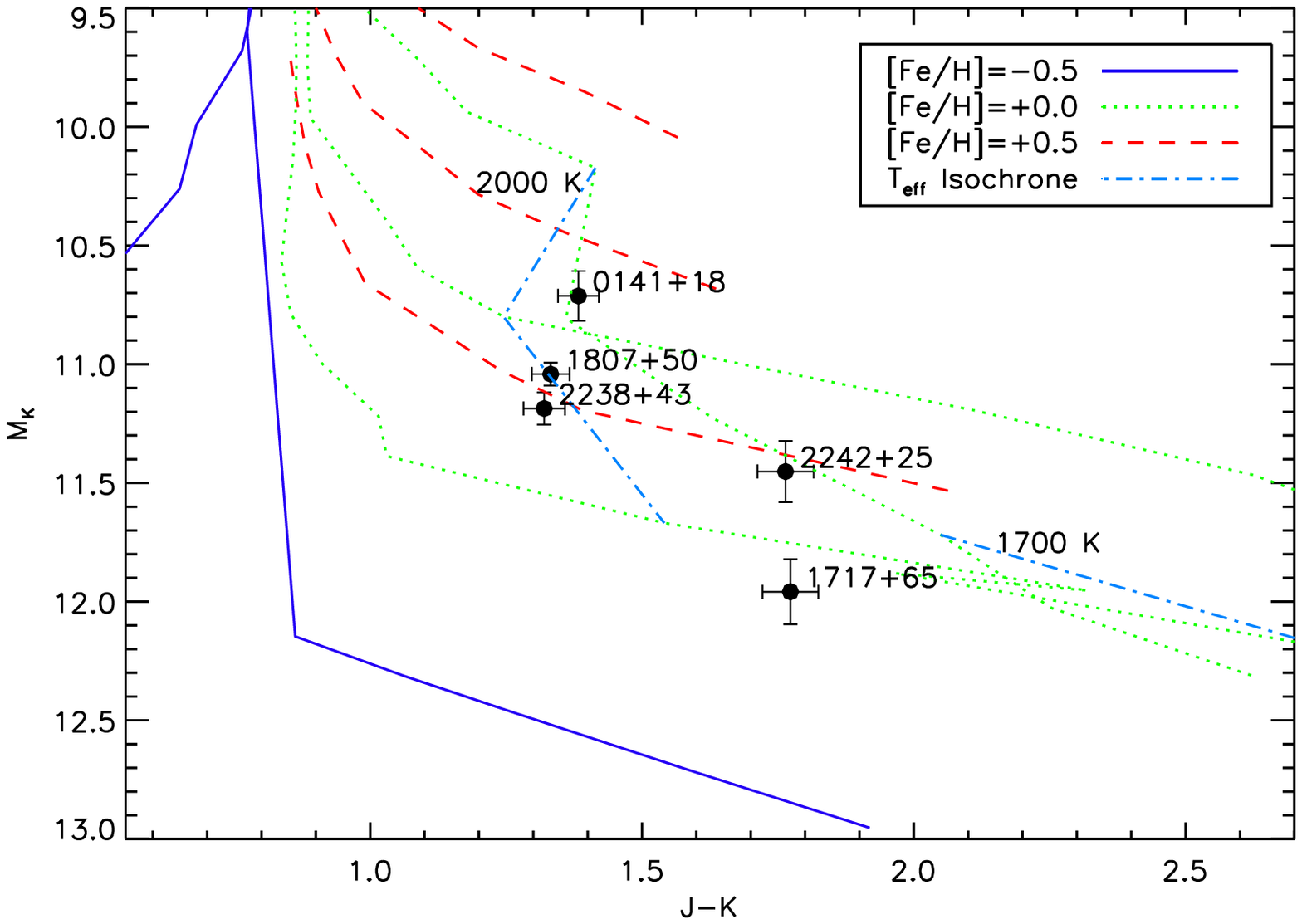}
\caption{Color-$M_{\rm K}$ diagrams of our five L dwarfs with models. The
  model tracks are the BSH06(left) and BT-Dusty(right) models. Different line-styles indicate
 different metallicities. For each metallicity, the 3 curves indicate different
 gravity. The gravity increases from bottom-left to top-right with values of
 log(g)=4.5, 5.0, 5.5. Thus, for a given mass evolutionary track, higher
 gravity models have fainter (larger) values of $M_{\rm K}$ and redder (larger)
 values of color. All magnitude are in {\rm 2MASS} system.}
\label{WPif6}
\end{figure}

In Fig. \ref{WPif6} we plot our objects on color versus absolute magnitude
diagrams. Model tracks from Burrows, Sudarsky \& Hubeny 2006 (hereafter BSH06)
and Allard et al. 2009 (http://phoenix.ens-lyon.fr/Grids/BT-Dusty/, hereafter
BT-Dusty) are overplotted for comparison. The BSH06 model grids cover log(g)
of 4.5, 5.0, and 5.5 (gravities in {\rm cgs}) and effective temperatures from
700 K to 2200 K, with metallicity of [Fe/H] = {-0.5, 0, +0.5}. The BT-Dusty
model grids cover log(g) of 4.5, 5.0, 5.5 and effective temperatures from 1500
to 3500 {\rm K}, with metallicities of [Fe/H] = {-0.5, 0.0, +0.5}. The
synthetic colors and absolute magnitudes are derived convolving the model
spectra with the {\rm 2MASS} filter profiles \citep[see][]{mar10}. The BSH06
model grids supply the flux at the surface of the object and at 10 pc. The latter
calculation assumes the radius-log(g)-$T_{\rm eff}$ relation from
\citet{bur97}. The BT-Dusty model grids only provide the flux at the surface of the
object, so, to calculate the absolute magnitudes, we calculate the radius
associated with each model spectrum by interpolating the BT-Dusty isochrones.

From Fig. \ref{WPif6} we can see the BSH06 and BT-Dusty can fit the colors
of these L dwarfs. In principle, we could determine metallicity or gravity
information from them.  But because of the known degeneracy between gravity
and metallicity, our objects can be described by different combinations of
the two parameters. This prevents assignment of a single gravity or
metallicity based only on this diagram. Nonetheless, our targets can be fitted by
solar or higher metallicity with log(g) between 5.0 and 5.5.  This is
consistent with the thin disk membership found in Section 4.2.  We note that
SD1717+6526 seems located outside the BT-Dusty model tracks. SD1717+6526 is an
L4, so we tentatively conclude that BSH06 and BT-Dusty are more consistent at
high temperatures ($\sim$2100~K) than at lower temperatures ($\sim$
1700~K). We will investigate this further in the Section 5.3.

\section{Temperatures and Luminosities}

In this section we derive effective temperature, bolometric luminosity,
radius, gravity and mass using the Chabrier et al.  (2000, hereafter CBA00)
dusty evolutionary model.  Then we combine observational spectra with
synthetic spectra from two brown dwarf models(BSH06 or BT-Dusty respectively)
 and the Stefan-Boltzmann law to estimate temperatures and luminosities.

\subsection{Physical parameters from an evolutionary model}

\begin{figure}[!htp]
\includegraphics[width=3.0in]{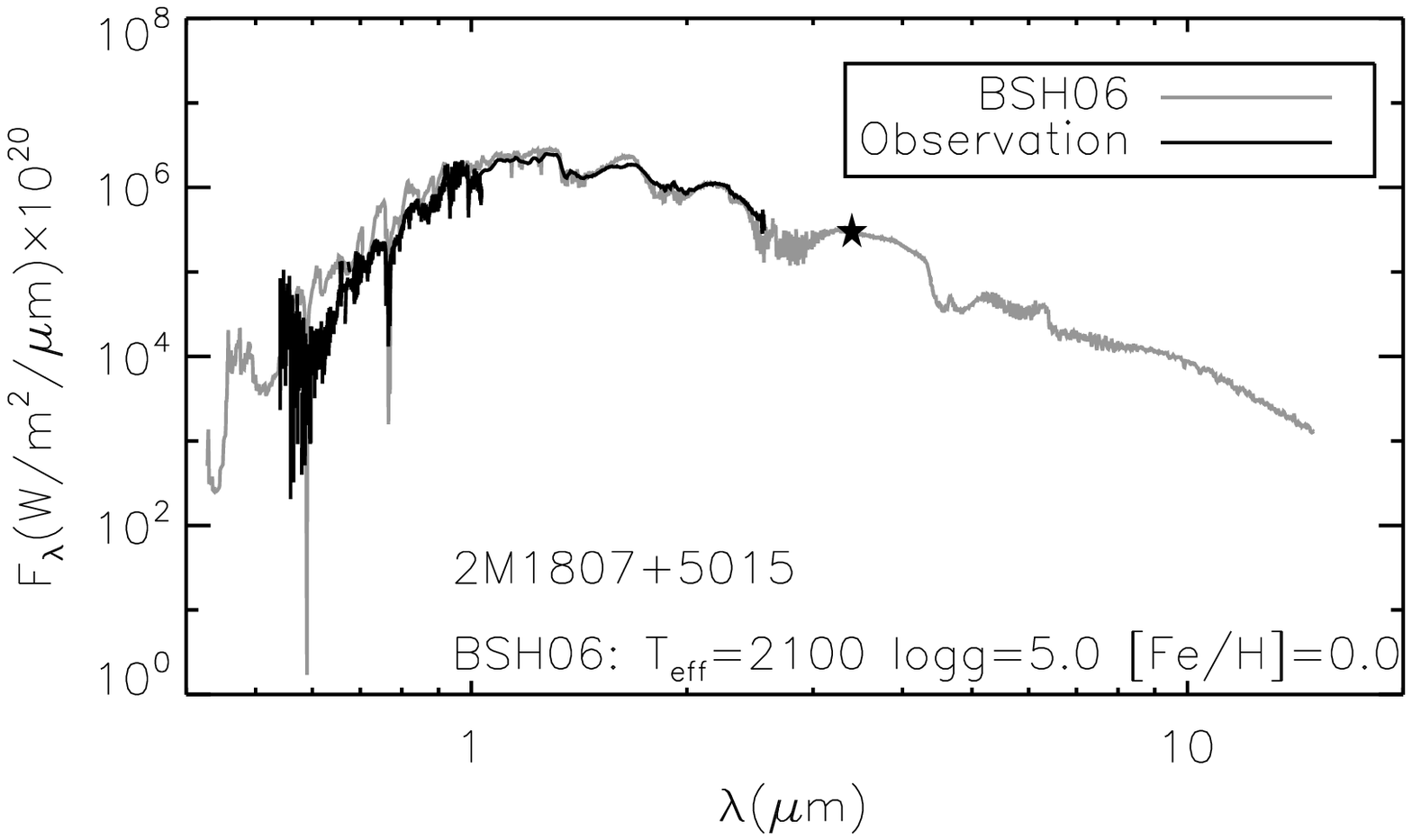}
\includegraphics[width=3.0in]{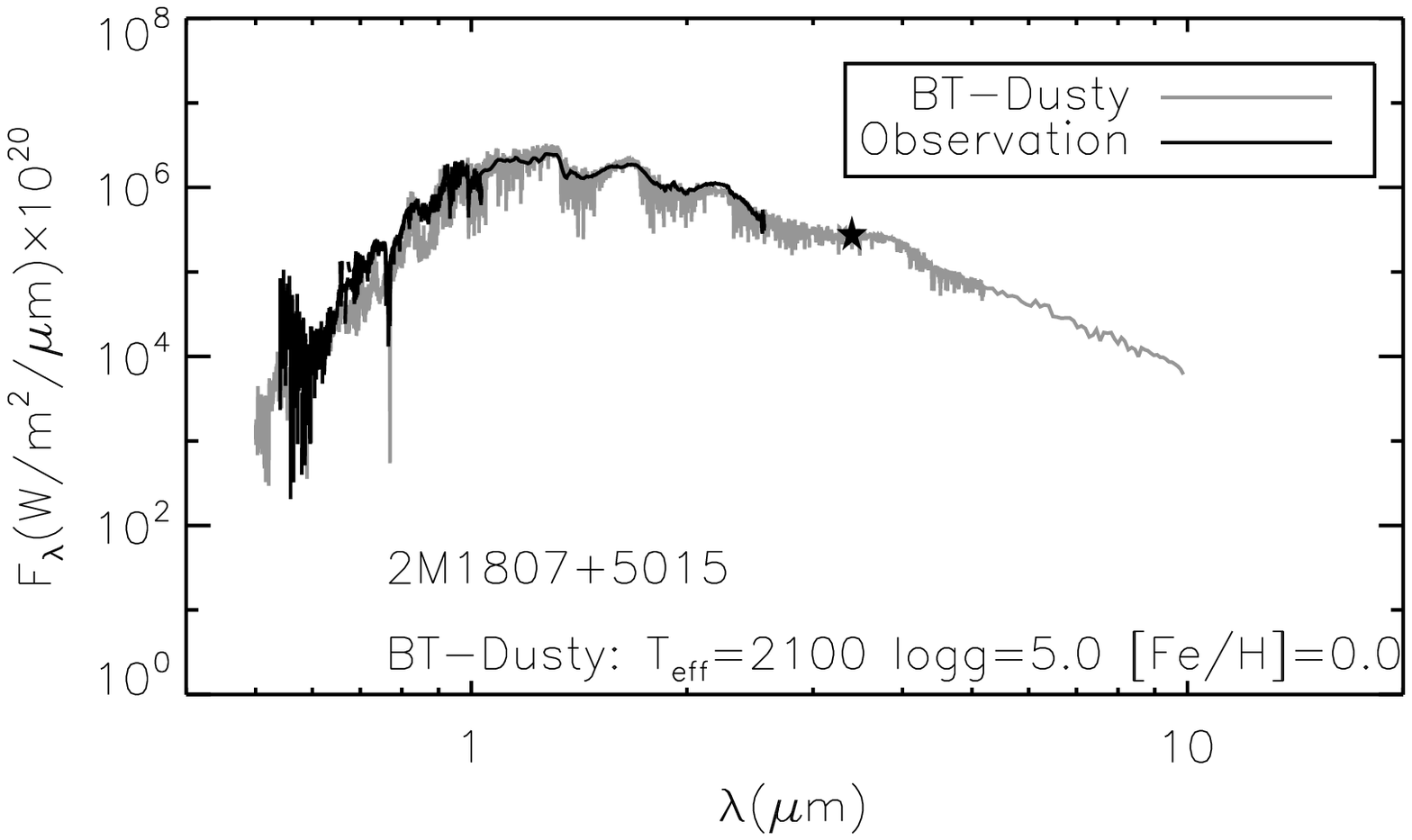}
 \caption{Left: absolute magnitude-effective temperature diagram. Right:
  luminosity-radius diagram. The two diagrams are plotted according to the CBA00
   dusty evolutionary model assuming ages of our targets between 0.5 and 10 Gyr.
   For this age range, the targets' radius change only $\sim$ 5\%.
    Interpolating the derived M$_K$ or luminosity we can find the effective
temperature or radius.}
\label{WPif7}
\end{figure}

We directly found the effective temperature and other parameters for all our
targets using the CBA00 dusty evolutional models and our derived parallax. The
gravity of our targets are between log(g)=5.0 and 5.5 (see Section
4.3). Given that these objects have higher gravity than young field dwarfs
\citep[e.g.][]{cru09}, and following our findings in Section 4.2, we assume
our targets to be between 0.5 and 10 Gyr old. For this age range the dusty
evolutionary model CBA00 provides relations between M$_K$ and effective
temperature, radius, bolometric luminosity, gravity and mass
(Fig. \ref{WPif7}).  Combining our parallax with 2MASS magnitudes we obtain
M$_{\rm K}$ and using CBA00 we find the parameters listed in Table
\ref{WPit8}.

\subsection{Temperatures and Luminosities from the Stefan-Boltzmann law}

We have obtained both optical \citep[from][]{cru07} and infrared
\citep[from][]{cru13,bur08b,bur10} spectra for three of our targets: 2M0141+1804,
2M1807+5015 and 2M2242+2542. To calibrate the spectra in flux, we used the $z$
band magnitude (for 2M1807+5015 we use z$_{\rm est}$ in Table \ref{WPit1}) in
the optical, and 2MASS J band photometry in the near-infrared. To calculate
the bolometric flux, we combined the observational spectra with the BSH06 and the
BT-Dusty models. We calibrated the flux level of the model spectra using WISE
W1 magnitudes, since these are well calibrated long wavelength measurements
and allow the spectra to join reasonably with the observed K band.  To
calculate an effective temperature range, following a similar method to
\citet{mar10}, we use the classical Stefan-Boltzmann law and the relationship
between $F_{\rm bol}$, $L_{\rm bol}$ and $T_{\rm eff}$

\begin{equation}
F_{\rm bol} = L_{\rm bol}/4\pi D^2,
\hspace{3mm} L_{\rm bol} = 4 \pi \sigma R^2 T_{\rm eff}^4
\label{WPie2}
\end{equation}

Integrating the observed optical and near-infrared spectra we obtained a preliminary
flux, which combined with our parallax yields a luminosity. Interpolating the
CBA00 luminosity-radius relationship (see Fig. \ref {WPif7}), we derived
the model predicted radius for our targets. Having the radius and
preliminary flux, we then obtained a preliminary effective
temperature. For the moment we do not consider metallicity and gravity,
this is discussed below. Using this temperature, we can choose the
appropriate model spectra. We then integrated the spectral energy distribution
(formed by optical, near-infrared and model spectra) to recalculate the bolometric flux, and
therefore a more precise temperature. Iterating the above procedure twice, we
obtain the bolometric flux, luminosity and effective temperature listed in
Table \ref{WPit9}.

When we choose the model spectra for an object, for either BSH06 or BT-Dusty model
spectra, we assume the targets have solar metallicity, and test two values of gravity:
 log(g)=5.5 and log(g)=5.0. Usually there are four models available,
taking 2M1807+5015 for example, the preliminary temperature is between
1875 and 1985 K (see Table \ref{WPit9}), and we choose the four synthetic
spectra with closest model parameters amongst the grids available: in this case
1900K, log(g)=5.5; 1900K, log(g)=5.0; 2000K,
log(g)=5.5; 2000K, log(g)=5.0. For each of the chosen model spectra we then
overlap with our observational spectra in order to create a full energy distribution.
We then output $F_{\rm bol}$, $L_{\rm bol}$, $T_{\rm eff}$ values.
The smallest and largest values generated by this process enable us to find
the range for each parameter given in columns 5-10 of Table \ref{WPit9}.
We note that the model grids available offer one or two synthetic spectra for
each temperature which correspond to different values of log(g). Thus we are
not in a position to estimate reliable gravities for our targets.

The uncertainty on temperature is calculated via standard propagation of the
errors on flux, distance and radius. For 2M2242+2542,
the flux errors in the optical and near-infrared bands were derived using the
average flux errors of 2M0141+1804 and 2M1807+5015 since they are not available
in the spectral file. The model spectra files do not provide
the errors on flux. However, when calculating the effective temperature, we used
BSH06 and BT-Dusty model grids to locate the appropriate synthetic spectra for
 each object and so find the uncertainty on the synthetic spectra flux. So that
  we can calculate the uncertainty on bolometric flux. To 
determine the error on radius we use the spread between the two values derived
from the CBA00 evolutionary model (see Table \ref{WPit9}). The final temperature
errors obtained are listed in columns 7 and 10 of Table
\ref{WPit9}. We note that the uncertainties on temperatures reflected the ranges of
the temperatures, which are dominated by the radius errors, \citet{bur08a} find
the same conclusion with a similar approach although they
did a piecewise scaling of the model spectra using multi-band photometry.

\subsection{Comparison}

A comparison of the temperatures in Tables \ref{WPit8} and \ref{WPit9}
indicates that they are consistent for each individual object. Effective
temperatures using the BSH06 and BT-Dusty models are very close for
2M0141+1804 and 2M1807+5015 with differences of $\sim$20 {\rm K}. The three
temperatures for 2M2242+2542 are consistent, but the differences between them
are larger, which is also reflected by the large errors on temperature (see
Table \ref{WPit9}). The temperature from the BT-Dusty model is slightly higher
than that from BSH06, which is close to the CBA00 one.

We should note that 2M2242+2542 is of spectral type L3, and is therefore
clearly cooler and with different features from the other L1 targets. The flux
of the BT-Dusty models does not join well with the observed spectrum,
over-predicting the flux level at K band. We therefore expect the model flux
to be higher than the object one in the mid-infrared as well, leading to an
over-estimation of its $T_{\rm eff}$. We would conclude that for this object
the fit given by the BT-Dusty models is not as accurate as the BSH06 one.

\section{Comments on individual targets}

{\bf 2M0141+1804:} Our results on temperature are consistent with Sengupta \&
Marley (2010) who estimate a temperature of 1850 $\pm$ 250 {\rm K} using
equations 3 and 4 of \citet{ste09}. The large error of 250 { \rm K} is due to
the difference in optical and IR spectral types and the authors used both
values when calculating the temperature. Our results have a smaller range in
temperature, because we have the optical and near-infrared spectra, which allow us to
get a relatively precise luminosity hence effective temperature. The radial
velocity is 24.7 km/s reported by \citet{bla10}. Our proper motions are
within one sigma of those in Casewell et al. (2008) though ours are significantly
more precise.

{\bf SD1717+6526:} The proper motions are consistent with those of
\cite{fah09} and the photometric distance we calculated based on Dupuy \& Liu
(2012) is within one standard deviation of the trigonometric distance.

{\bf 2M1807+5015:} This object is the brightest of our five targets.
\citet{sen10} reported a T$_{eff}$=2100$\pm$100 {\rm K} and \citet{wit11}
derived T$_{eff}$=1900 {\rm K}, log{\em g}=5.5, [Fe/H]=0.0 through
drift-phoenix model fitting. Our results are more consistent with the lower
value.  \citet{sei10} reported a radial velocity of -0.4 km/s and very low
values for the U,V,W velocity components, which is consistent with our
results.

{\bf 2M2238+4353:} \citet{ber10} reported it is a binary candidate with a mass
ratio of 0.57-0.84 assuming an age between 1 and 5 Gyr. However we do not
see any binary signature in our parallax determination residuals. Also, the
position in Fig. 4 does not indicate binarity, though, unless the mass ratio
is larger than about 0.6, we would not expect to see any significant
brightening.

{\bf 2M2242+2542:} \citet{bou03} observed this object using Hubble Space
Telescope in a search for binaries and concluded it was a single object, which
is consistent with our conclusions. \citet{giz03} and \citet{cru07} derived
photometric distances of $\sim$ 30 and $\sim$ 27 pc respectively. We find a
trigonometric distance of $\sim$21 pc which is consistent with the photometric
distance from Table \ref{WPit3}.

\section{Summary and future work}

We report the parallaxes and proper motions of five L dwarfs using a robotic telescope. Our
trigonometric distances are very close to the photometric ones. Our proper
motions are consistent with the literature but have smaller
errors. Examinations of the objects' spectral type versus absolute magnitude,
U versus V velocity and color versus absolute magnitude over-plotted with
model tracks diagrams indicate that the five L dwarfs are single thin disk
objects with solar metallicity.  For all five objects, effective temperature,
luminosity and bolometric flux, radius, gravity, and mass are derived from the
CBA00 model.  For three of our targets we derived the effective temperature
combining their measured spectra with atmospheric models (BSH06 and BT-Dusty)
to determine the bolometric flux.  We found current low mass models do not
work well at lower temperatures compared to higher temperatures. We find that
BSH06 and CBA00 predict more consistent temperatures for the lower temperature
objects than the BT-Dusty model but we also note our sample size is small and
the error on T$_{eff}$ is large when using BSH06. Further model testing with a
bigger sample is needed to see if these effects are real.

This work is the first parallax determination using a ground-based robotic
telescope. Parallax determinations have stringent observational requirements
which are efficiently satisfied by robotic scheduling. The requirement for
long term stability and repeatability is also well met by robotic
procedures. The RATCam camera is scheduled to be completely decommissioned in
2013 and be replaced by the Infrared-Optical (IO) camera though RATCam and IO
(with only a z-band filter) are both working currently.  Once IO is fully
commissioned, and, with the lessons learnt from this programme, we plan to
launch a more ambitious programme to observe the nearby and rapidly expanding
sample of interesting L and T dwarfs which are available for the Liverpool
Telescope.

\section*{Acknowledgments}
We thank Leigh Smith from the University of Hertfordshire for his help in
analyzing the image data. We thank Zhenghong Tang, Yong Yu and Zhaoxiang Qi
from Shanghai Astronomical Observatory for their helpful discussion on the
centroiding precision. This work is partially funded by IPERCOOL n.247593
International Research Staff Exchange Scheme and PARSEC n.236735 International
Incoming Fellowship within the Marie Curie 7th European Community Framework
Programme. YW and ZS acknowledge the support of NSFC10973028, 10833005, 10878003 and
NKBRSF2007CB815403.  RS and HRAJ acknowledge the support of Royal Society
International Joint Project 2007/R3.  This research has benefited from the M,
L, and T dwarf compendium housed at DwarfArchive.org and maintained by Chris
Gelino, Davy Kirkpatrick, and Adam Burgasser. The Liverpool Telescope is
operated on the island of La Palma by Liverpool John Moores University in the
Spanish Observatorio del Roque de los Muchachos of the Instituto de
Astrofisica de Canarias with financial support from the UK Science and
Technology Facilities Council.

\begin{deluxetable}{llllrrrrlll}
\tabletypesize{\scriptsize}
\tablecaption{Magnitudes and spectral types of the targets.}
\tablewidth{0pt}
\tablehead{
\colhead{Short Name} & \colhead{Discovery Name} &\colhead{$z$$_{\rm SDSS}$} & \colhead{$z_{\rm est}$} &
\colhead{J$_{\rm 2MASS}$}& \colhead{H$_{\rm 2MASS}$}&\colhead{K$_{\rm 2MASS}$} & \colhead{{\rm W1}}& \colhead{SpT$_{\rm opt}$} & \colhead{SpT$_{\rm NIR}$}
}
\startdata
 2M0141+1804 & 2MASS J0141032+180450 & 16.34 & {\em 16.34} &13.88 & 13.03 & 12.49 & 12.16 & L1$^1$  & L4.5$^3$ \\
 SD1717+6526 & SDSS J171714.10+652622.2 & 17.79 & {\em 17.67} & 14.95 & 13.84 & 13.18 & 12.53 & L4$^2$ &  - \\
 2M1807+5015 & 2MASSI J1807159+501531 & - & {\em 15.43} & 12.93 & 12.13 & 11.60 & 11.25 & L1.5$^4$ & L1$^3$ \\
 2M2238+4353 & 2MASSI J2238074+435317 & - & {\em 16.42} & 13.84 & 13.05 & 12.52  & 12.20 & L1.5$^4$ &  - \\
 2M2242+2542 & 2MASS J22425317+2542573& 17.49 & {\em 17.42} & 14.81 & 13.74 & 13.05 & 12.51 & L3$^1$ & L$^1$ \\
\enddata

\label{WPit1}
\tablenotetext{a}{z$_{\rm est}$ is an z-band magnitude estimated from the z-J
  color -  optical spectral type relation from
  \citet{zha09}. SpT$_{\rm opt}$ is the spectral type obtained from optical spectra
  and SpT$_{\rm NIR}$ from the near infrared spectra.  \\
  References. $^1$\citet {cru07}; $^2$\citet {haw02}; $^3$ \citet{wil03};
  $^4$\citet {cru03}; $^5$\citet{zha10}.}
\end{deluxetable}

\begin{deluxetable}{lllllllllr}
\tabletypesize{\scriptsize}
\tablecaption{Parallaxes and proper motions derived for our targets.}
\tablewidth{0pt}
\tablehead{
 \colhead{Short Name}& \colhead{RA,Dec} & \colhead{Epoch}&\colhead{$N_{o},N_{r}$} & \colhead{$\Delta$t}& \colhead{$\pi$}& \colhead{COR}  & \colhead{$\mu_{\alpha}cos\delta$} & \colhead{$\mu_{\delta}$} & \colhead{V$_{tan}$}\\
\colhead{}&\colhead{(hh mm ss),(dd mm ss)}&\colhead{}&\colhead{} &\colhead{(yrs)}&\colhead{(mas)}&\colhead{(mas)}& \colhead{(mas/yr)}& \colhead{(mas/yr)}           & \colhead{(km/s)}
}
\startdata

2M0141+1804&01:41:03.5,+18:04:49.5&2008.66&40,9 & 4.35  & 44.06$\pm$2.05 & 1.77 & 405.2$\pm$1.1 & -48.7$\pm$0.9 &43.9$\pm$2.0\\
2M1717+6526&17:17:14.2,+65:26:21.2&2008.60&65,5 & 7.62  & 57.05$\pm$3.51 & 1.46 & 150.2$\pm$1.0 & -109.3$\pm$0.6 &15.4$\pm$1.0\\
2M1807+5015&18:07:15.9,+50:15:30.2&2009.27& 57,18& 7.04  & 77.25$\pm$1.48 & 1.68 &  27.2$\pm$1.0 & -130.2$\pm$1.5 &8.1$\pm$0.2\\
2M2238+4353&22:38:07.7,+43:53:16.6& 2009.49&52,37& 7.71  & 54.11$\pm$1.55 & 1.24 & 324.3$\pm$0.5 & -121.0$\pm$0.4 &30.3$\pm$0.9\\
2M2242+2542&22:42:53.4,+25:42:56.6& 2009.53&53,11 & 7.83 & 47.95$\pm$2.74 & 1.71 & 382.0$\pm$0.9 &  -64.6$\pm$0.7 &38.3$\pm$2.2\\
\enddata

\label{WPit2}
\tablenotetext{a}{The columns denote object name, right ascension (RA) and
  declination (Dec) of the base frame, epoch of the base frame,
number of frames and number of reference objects($N_{o},N_{r}$), total time span for observations ($\Delta$t), absolute parallax ($\pi$), correction from relative to absolute parallax (COR), proper motions in RA($\mu_{\alpha}cos\delta$), proper motion in Dec($\mu_{\delta}$) and tangential velocity(V$_{tan}$).}
\end{deluxetable}

\begin{deluxetable}{lrrrrr}
\tabletypesize{\scriptsize}
\tablecaption{Photometric and trigonometric distances of the five L dwarfs.}
\tablewidth{0pt}
\tablehead{
 \colhead{Short Name}&  \colhead{$D_J$} &  \colhead{$D_H$} &  \colhead{$D_K$}& \colhead{$<D_P>$} &\colhead{$D_{\pi}$}\\
 \colhead{}&  \colhead{(pc)}& \colhead{(pc)}& \colhead{(pc)}&\colhead{(pc)}& \colhead{(pc)}
 }
 \startdata
 2M0141+1804& 24.2$\pm$4.2 & 24.2$\pm$4.2 & 22.6$\pm$3.9&23.7$\pm$4.1& 22.7$\pm$1.1 \\
 SD1717+6526& 22.9$\pm$3.4 & 22.0$\pm$3.3 & 20.3$\pm$3.2&21.7$\pm$3.3& 17.5$\pm$1.7  \\
 2M1807+5015& 14.4$\pm$2.5 & 14.9$\pm$2.6 & 14.0$\pm$2.4&14.4$\pm$2.5& 12.9$\pm$0.3 \\
 2M2238+4353& 21.9$\pm$3.8 & 22.7$\pm$4.0 & 21.3$\pm$3.6&22.0$\pm$3.8& 18.5$\pm$0.6 \\
 2M2242+2542& 26.1$\pm$3.9 & 24.8$\pm$3.7 & 21.9$\pm$3.4&24.2$\pm$3.7& 20.9$\pm$1.2 \\
\enddata
\label{WPit3}
 \tablenotetext{a} {We calculated the spectrophotometric distances according to
  the J,H and K band SpT - absolute magnitude relationship of Dupuy \& Liu (2012).
   $<D_P>$ is the weighted mean spectrophotometric
   distance and $D_{\pi}$ is the distances derived from our trigonometric parallax.}
\end{deluxetable}

\begin{deluxetable}{lcc}
\tabletypesize{\scriptsize}
\tablecaption{Comparison of our proper motions with literature values.}
\tablewidth{0pt}
\tablehead{
 \colhead{Short Name}& \colhead{Table 2 $\mu_{\alpha}cos\delta$,$\mu_{\delta}$}  &  \colhead{Literature $\mu_{\alpha}cos\delta$,$\mu_{\delta}$}\\
 \colhead{}    &  \colhead{(mas/yr)}                   &   \colhead{(mas/yr)}
 }
 \startdata
2M0141+1804& 405.2$\pm$1.1, -48.7$\pm$0.9 & 425.1$\pm$17.6,-32.2$\pm$16.5$^1$  \\
SD1717+6526& 150.2$\pm$1.0, -109.3$\pm$0.6 & 159.0$\pm$7.0, -92.0$\pm$16.0$^3$\\
2M1807+5015&  27.2$\pm$1.0, -130.2$\pm$1.5 & 34.6$\pm$18.5,-125.7$\pm$14.3$^2$ \\
2M2238+4353& 324.3$\pm$0.5, -121.0$\pm$0.4& 324.0$\pm$12.0,132.0$\pm$16.0$^3$ \\
2M2242+2542& 382.0$\pm$0.9, -64.6$\pm$0.7 & 408.8$\pm$15.5,-45.2$\pm$16.2$^2$  \\
\enddata
\label{WPit4}
 \tablenotetext{a}{References. {$^1$}\citet{cas08}, {$^2$}\citet{jam08},{$^3$}\citet{fah09}}
\end{deluxetable}

\begin{deluxetable}{lllllllll}
\tabletypesize{\scriptsize}
\tablewidth{0pt}
\tablecaption{Coefficients of equation \ref{WPie1} fitting objects in figure \ref{WPif4} excluding known and possible binaries.}
\tablehead{
 \colhead{Mag.}& \colhead{a$_0$} & \colhead{a$_1$}& \colhead{a$_2$}& \colhead{a$_3$}& \colhead{a$_4$}& \colhead{a$_5$}&\colhead{a$_6$}& \colhead{$\sigma$}
 }
 \startdata
M$_J$& -1.33315e2 & 5.33338e1& -7.99504& 6.21422e-1& -2.6167e-2&  5.64852e-4& -4.88497e-6&0.456012 \\
M$_H$& -8.62209e1 & 3.48403e1& -5.12332 & 3.93049e-1& -1.64117e-2&3.53139e-4& -3.06005e-6& 0.400478 \\
M$_K$& -1.24473e2 & 4.98554e1& -7.50172 & 5.85077e-1&  -2.47521e-2 & 5.37504e-4& -4.67296e-6& 0.403721 \\
\enddata
\label{WPit5}
\end{deluxetable}

\begin{deluxetable}{lrrrr}
\tabletypesize{\scriptsize}
\tablewidth{0pt}
\tablecaption{Calculated $U, V$ and $W$ for our five targets.}
\tablehead{
\colhead{Short Name}& \colhead{$V_{rad}$} & \colhead{$U$}& \colhead{$V$} & \colhead{$W$}\\
\colhead{}& \colhead{km/s} & \colhead{km/s}& \colhead{km/s} &\colhead{km/s}
}
\startdata
2M0141+1804& 24.7 & -56.6  & -6.2&  -5.0  \\ \hline
\multirow{3}{*}{SD1717+6526}&30& -3.1  &44.0 &15.2 \\
 &-30.0 & 1.6 &-5.3 & -18.6 \\
 &  0.0 & -0.8 &19.4 &-1.7 \\ \hline
2M1807+5015& -0.4 & -3.5& 11.7 & 4.4 \\ \hline
\multirow{3}{*}{2M2238+4353}& 30.0 & -33.9 &   33.0 &-22.3 \\
&-30.0 & -24.8 &-24.8 &-8.9\\
&  0.0 & -29.3 &4.1 &-15.6 \\ \hline
\multirow{3}{*}{2M2242+2542}& 30.0 & -39.4 &26.2 &-29.8 \\
&-30.0 & -39.6&-26.4 &-1.0 \\
&  0.0 & -39.5 & -0.1 &-15.4\\
\enddata
\label{WPit6}
\tablenotetext{a}{Note. We assume the three L dwarfs without measured radial velocities to be
0 and $\pm$ 30 km/s as radial velocities following the SDSS M dwarfs' distribution. }
\end{deluxetable}

\begin{deluxetable}{llll}
\tabletypesize{\scriptsize}
\tablewidth{0pt}
\tablecaption{"Critical" radial velocities and halo probabilities.}
\tablehead{
\colhead{Short Name}& \colhead{$V_{rad1}$}& \colhead{$V_{rad2}$}& \colhead{P}\\
\colhead{} & \colhead{km/s} & \colhead{km/s}& \colhead{\% }
 }
 \startdata
SD1717+6526& -179.1 &48.0& 5.5\\
2M2238+4353&-137.8& 49.2& 5.1\\
2M2242+2542& -136.9& 57.6 &2.8\\
\enddata
\label{WPit7}
\tablenotetext{a}{Note. These radial velocities for the
  three targets locate them on the 2 $\sigma$ ellipsoid in
  Fig. \ref{WPif5}. Integrating the Gaussian profile of the radial velocity
  outside the 2$\sigma$ velocities for each target we get the probability(P)
 that the targets are halo objects.}
\end{deluxetable}

\begin{deluxetable}{lllllll}
\tabletypesize{\scriptsize}
\tablewidth{0pt}
\tablecaption{Temperature, luminosity, radius, gravity and mass derived from CBA00 model.}
\tablehead{
\colhead{Short Name} & \colhead{$M_k$}& \colhead{T$_{eff}$}& \colhead{Luminosity} & \colhead{Radius} & \colhead{Gravity} & \colhead{Mass}\\
\colhead{}& \colhead{}& \colhead{({\rm K})}& \colhead{$log_{10} (L/L_\odot$)}& \colhead{$R/R_\odot$}& \colhead{log$_{10}$(g)}& \colhead{M/M$_\odot$}
}
\startdata
2M0141+1804&10.71&2225-2305  &-3.63-(-3.60) &0.1000-0.1055 &5.21-5.34&0.067-0.080\\
SD1717+6526 & 11.96&1450-1563 &-4.41-(-4.36) &0.0860-0.1040 &5.00-5.41&0.040-0.070\\
2M1807+5015&11.04&2000-2138  &-3.82-(-3.78) &0.0958-0.1036 &5.18-5.36&0.058-0.077\\
2M2238+4353& 11.19&1828-2038 &-3.95-(-3.85) &0.0930-0.1030 &5.15-5.37&0.053-0.075\\
2M2242+2542&11.45&1688-1850  &-4.14-(-4.08) &0.0905-0.1032 &5.08-5.39&0.048-0.073\\
\enddata
\label{WPit8}
\tablenotetext{a} {The range of values shown are
found based on assuming age range between of 0.5 and 10 Gyr. }
\end{deluxetable}

\begin{deluxetable}{llll|lll|lll}
\tabletypesize{\scriptsize}
\tablewidth{0pt}
\tablecaption{Bolometric flux, luminosity, effective temperature from combination of observational and model spectra.}
\tablehead{
\multicolumn{1}{c}{} &
\multicolumn{3}{c|}{Preliminary  Parameters} &
\multicolumn{3}{c|}{Combining BSH06} &
\multicolumn{3}{c}{Combining BT-Dusty} \\ \hline
\colhead{Short Name} &\colhead{L} & \colhead{\Rsun}  &  \colhead{$T_{\rm eff}$} &  \colhead{$F_{\rm bol}$} & \colhead{L$_{\rm bol}$} & \colhead{$T_{\rm eff}(\sigma)$}
 & \colhead{$F_{\rm bol}$} & \colhead{L$_{\rm bol}$} & \colhead{$T_{\rm eff}(\sigma)$}}
\startdata
 2M0141+1804& -3.80 &0.1038-0.0950& 2015-2107 & 1.24-1.24&-3.69&2126-2206(101)& 1.30-1.30 &-3.68 & 2147-2216(105)  \\
 2M1807+5015& -3.93 &0.1030-0.0919& 1875-1985 & 2.83-2.84&-3.83&1982-2071(68)& 2.97-2.98 & -3.81& 1999-2085(69)\\
 2M2242+2542& -4.24 &0.1035-0.0885& 1560-1687 & 0.618-0.624&-4.11&1699-1823(225)& 0.715-0.751 & -4.03& 1736-1844(133) \\
 \enddata
\tablenotetext{a}{Notes. Columns 2-4 are preliminary parameters: luminosity, radius in R$_\odot$, effective temperature ($T_{\rm eff}$).
 Columns 5-7 and Columns 8-10 are final parameters after two iterations combining BSH06 and BT-Dusty respectively:
bolometric flux, luminosity and temperature. Luminosity is in units of \Lsun, bolometric flux $F_{\rm bol}$ in units
of (x10$^{-14}{\rm J/m^2}$) and temperature in units of {\rm K}.}
\label{WPit9}
\end{deluxetable}

\newpage


\end{document}